\documentclass[a4paper,11pt]{article}
\usepackage{graphicx,subcaption}
\usepackage{subcaption}
\usepackage{caption}
\usepackage{pos}

\DeclareMathOperator{\tr}{Tr}

\newcommand{\dif}[1]{\mathrm{d} #1 }

\begin{document}
\title{Coarse graining in effective theories of lattice QCD in 1+1d and 2+1d}
\author*{Christoph Konrad}
\author{Owe Philipsen}
\affiliation{Institute for Theoretical Physics, Goethe-University Frankfurt am Main,\\Max-von-Laue-Str. 1,  60438 Frankfurt am Main, Germany}
\emailAdd{konrad@itp.uni-frankfurt.de}
\emailAdd{philipsen@itp.uni-frankfurt.de}
\abstract{In the strong coupling and heavy quark mass regime, lattice QCD dimensionally reduces to effective theories of 
	Polyakov loops depending on the parameters of the original Wilson action $\beta, \kappa$ and $N_\tau$. We apply coarse graining techniques to such theories in 1d and 2d, corresponding to lattice QCD at finite temperature
	and non-zero chemical potential in 1+1d and 2+1d, respectively. In 1d the method is applied to the effective theories up to $\mathcal{O}(\kappa^4)$. Using the transfer matrix, the recursion relations are solved analytically. The thermodynamic limit is taken for some observables. Afterwards, continuum extrapolation is performed numerically and results are discussed. In 2d the coarse graining method is applied in the pure gauge and static quark limit. Running couplings	are obtained and the fixed points of the transformations are discussed. Finally, the critical coupling of the deconfinement transition is determined in both limits. Agreement to about 12\% with Monte Carlo results of 2+1d Yang-Mills theory from the literature is observed.}
\FullConference{The 38th International Symposium on Lattice Field Theory, LATTICE2021 26th-30th July, 2021\\
Zoom/Gather@Massachusetts Institute of Technology\\
26 - 30 July 2021
}
\maketitle
\section{Introduction}
The study of lattice QCD (LQCD) at non-zero chemical potential is still a difficult task due to the sign problem. Therefore, dimensionally reduced effective theories of LQCD with standard Wilson fermions at non-zero chemical potential have been derived by integrating out spatial gauge link variables after a combined expansion in the inverse coupling $\beta$ and hopping parameter $\kappa$. The actions are expressed in terms of Polyakov loops and, thus, have significantly fewer degrees of freedom and a milder sign problem compared to the mother theory. Additionally, they can be interpreted as complex valued spin models  \cite{Langelage2011, Schoen2018, Neuman2015, Glesaaen2016}. Here, we test coarse graining techniques on dimensionally reduced 1d and 2d effective theories, representing LQCD in 1+1d and 2+1d, respectively. 

The running couplings of the effective theories in 1d are obtained up to order $\mathcal{O}(\kappa^4)$ for general number of colors $N_c$. Afterwards, the corresponding recursion relations are solved analytically resulting in a compact expression for the partition function in terms of the transfer matrix. This is used to calculate the pressure and baryon density at non-zero chemical potential. 

For the 2d effective theories the running couplings are obtained in the pure gauge and static quark limit, respectively. In both cases the fixed point corresponding to the deconfinement transition is determined analytically. In the pure gauge limit our results for the critical coupling $\beta_c$ agree with Monte-Carlo results of 2+1d Yang-Mills theory from the literature to about 12\%.
\section{The effective theories}
To derive the effective theories one starts with the Wilson lattice action and integrates out the Grassmann fields. Next, integrating over spatial link variables defines an effective action $S_\text{eff}$ \cite{Langelage2014},
\begin{align}
	Z &= \int[\dif{U_\mu}]e^{-S_G[U_\mu]}\det Q[U_\mu] = \int[\dif{U_0}]e^{-S_\text{eff}[U_0]}\;,\notag\\
	 S_\text{eff}[U_0]	& = -\log \int[\dif{U_i}] e^{-S_G[U_0,U_i]}\det Q[U_0,U_i] = S_\text{G, eff} + S_\text{stat} + S_\text{kin, eff}\;,
\end{align}
where $\det Q(U_\mu)$ is the quark determinant whereas $S_\text{G, eff}, S_\text{stat} $ and $ S_\text{kin, eff}$ are the effective gauge, static quark and effective kinetic quark action, respectively. Gauge invariance then implies that $S_\text{eff}$ is a functional of Polyakov loops $S_\text{eff}[U_0] = S_\text{eff}[U]$ with $U_{\vec{x}}:=\prod_{\tau =0}^{N_\tau-1} U_0(\vec{x}, \tau)$ \cite{Langelage2011}.
\subsection{The effective gauge action}
The strong coupling expansion is implemented by a character expansion of the plaquette action~ \cite{Wipf2013}. In 1+1d the exact solution of pure gauge theory has been known for many decades \cite{Migdal1996}. In a similar manner, the 1d effective gauge action can be derived without further approximations. Then, the spatial link integrals can be performed. The effective gauge action is given by \cite{Langelage2011}
\begin{align}
	S_\text{G, eff} = -\sum_{\langle \vec{x},\vec{y}\rangle}\log\sum_r \lambda_r\chi_r(U_{\vec{x}}^\dagger)\chi_{r}(U_{\vec{y}})\;,\label{sEffGauge}
\end{align}
with $\chi_r$ being the character of the irreducible representation $r$. The couplings $\lambda_r$ are expressed in terms of the fundamental coefficients of the character expansion $u(\beta)$.  In higher dimensions the couplings $\lambda_r$ obtain additional corrections in terms of the $u(\beta)$ and long-range interactions appear in $S_\text{G, eff}$ \cite{Langelage2011}. Taking kinetic quarks into account leads to similar corrections in $\kappa$ to the couplings and results in new interaction terms. When considering higher dimensions or including kinetic quarks, $S_\text{G, eff}$ is usually truncated to include only the few lowest dimensional representations. In the simplest non-trivial case only the fundamental representation is considered \cite{Fromm2012},
\begin{align}
	S_\text{G, eff} = -\sum_{\langle \vec{x},\vec{y}\rangle} \log\left(1+\lambda_f L_{\vec{x}}L_{\vec{y}}^*+\lambda_f L_{\vec{y}}L_{\vec{x}}^*\right)\quad \text{with}\quad \chi_f(U_{\vec{x}}) =: L_{\vec{x}}\;.
\end{align}
\subsection{The effective quark action}
To derive the effective quark action, $\det Q$ is split into the static determinant $\det Q_\text{stat}$ and the kinetic quark determinant $\det Q_\text{kin}$ according to $\det Q  = \det Q_\text{stat} \det Q_\text{kin} $. The former can be evaluated exactly for $N_f$ degenerate quark flavors \cite{Langelage2014},
\begin{align}
	\det Q_\text{stat} = \prod_{\vec{x}} \det\left(1+h_1 U_{\vec{x}}\vphantom{U_{\vec{x}}^\dagger}\right)^{2N_f}\det\left(1+\bar h_1 U_{\vec{x}}^\dagger\right)^{2N_f} := \prod_{\vec{x}}\det Q_\text{stat}^\text{loc}(U_{\vec{x}}) = e^{-S_\text{stat}}\;,
\end{align}
with $h_1 = (2\kappa)^{N_\tau} e^{N_\tau a \mu}$ and $\bar h_1 = (2\kappa)^{N_\tau} e^{-N_\tau a \mu}$. 

The kinetic quark determinant has to be expanded in $\kappa$ to perform the integration over the spatial link integrals. Usually, this is done using the trace-log identity $\det(\cdot) = \exp \tr\log( \cdot)$ as well as expanding the exponential and logarithm to the desired order. After applying resummation techniques an effective kinetic quark action $S_\text{kin, eff}$ is obtained. It is expressed in terms of rational functions of temporal Wilson lines,
\begin{align}
	W_{nm\bar n \bar m}(U_{\vec{x}}) = \tr \frac{(h_1 U_{\vec{x}})^m}{(1+h_1 U_{\vec{x}})^n}\frac{(\bar h_1 U^\dagger_{\vec{x}})^{\bar m}}{(1+\bar h_1 U^\dagger_{\vec{x}})^{\bar n}}\;, \quad W^\pm_{nm\bar n \bar m} = W_{nm00}\pm W_{00\bar n \bar m}\label{fractionalWilsonLoop}\;,
\end{align}
 which can also be represented in terms of the Polyakov loop \cite{Langelage2014, Neuman2015, Glesaaen2016}. To leading order in $\kappa$ the effective kinetic quark action $S_\text{kin, eff}$ reads \cite{Fromm2012, Neuman2015, Glesaaen2016}
 \begin{align}
 	S_\text{kin, eff} = 2 h_2\sum_{\langle \vec{x},\vec{y}\rangle} W_{1111}^-(U_{\vec{x}})W_{1111}^-(U_{\vec{y}})\quad \text{with}\quad h_2 = N_f N_\tau \kappa^2 /N_c\;.
 \end{align}
 
 At $\mathcal{O}(\kappa^{2n})$ the non-local behavior of $\det Q_\text{kin}$ implies long-range and multipoint interaction terms over a taxi driver distance $n$ in $S_\text{kin, eff}$ \cite{Neuman2015, Glesaaen2016}. In 1d \cite{Schoen2018} and 3d \cite{Neuman2015} the effective action is known to $\mathcal{O}(\kappa^4)$, whereas in the cold and dense limit in 3d it has been derived to $\mathcal{O}(\kappa^8)$ \cite{Glesaaen2016}. 
 
 In the remaining parts of this work vector arrows are not written explicitly anymore.  
\section{Coarse graining of the 1d effective theories}
The effective action is a functional of traced powers of Polyakov loops living on the spatial coordinates of the lattice. We can interpret those as continuous and complex-valued spin variables~ \cite{Langelage2011}. This suggests the use of renormalization schemes that have previously been applied to Ising-type systems. For the 1d effective theory with only nearest neighbor interactions this corresponds to integrating out every second lattice site. By contrast, next-to-nearest neighbor interactions require integrating out every second pair of lattice sites \cite{Wipf2013}. 
\subsection{The static quark limit}
Because the static determinant depends only on temporal links, the partition function for static quarks is simply
determined by including $\det Q_\text{stat}$ into the integrand and no truncation of the sum over irreducible representations is necessary. Typically, a renormalized action includes all possible combinations of field interactions. Thus, we choose the recursion approach
\begin{align}
	Z = \int [\dif{U_x}]^{(n)} \det Q_\text{stat}^{(n)} \prod_{x=1}^{N_s^{(n)}} \sum_{r,r'}\lambda_{rr'}^{(n)} \chi_r(U_x) \chi_{r'}(U_{x+1}^\dagger)\quad\text{with} \quad \lambda_{rr'}^{(0)} = \lambda_r \delta_{rr'}\;.
\end{align}
After iterating the renormalization procedure once, we find the running couplings,
\begin{align}
	\lambda_{r_1^{} r_2^{}}^{(n+1)} = \sum_{r_1^{\prime},r_2^{\prime}}\lambda_{r_1^{} r_1^{\prime}}^{(n)}\lambda_{r_2^{\prime} r_2^{}}^{(n)} \int\dif{U} \det Q_\text{stat}^\text{loc}(U) \chi_{r_1^{\prime}}(U^\dagger) \chi_{r_2^{\prime}}(U)\label{staticQuarksRunningCoupling}\;.
\end{align}
In the pure gauge limit $\det Q_\text{stat}^\text{loc} = 1$. Thus the gauge integral in \eqref{staticQuarksRunningCoupling} can be solved with the orthogonality relation for irreducible characters. Then one finds the solution to the renormalization scheme,
\begin{align}
	\lambda_{rr'}^{(n)} = \delta_{rr'}\lambda_r^{2^n}\;, \quad Z = \sum_r \lambda_r^{N_s}\;,
\end{align}
in agreement with the exact solution of 1+1d pure gauge theory, if $N_s$ is a power of two \cite{Migdal1996, Wipf2013}.
\subsection{Next-to-leading order corrections}
In the next scenario we include kinetic quarks to $\mathcal{O}(\kappa^4)$ in our theory. Taking into account the next-to-nearest neighbor interactions, we integrate out every second pair of lattice sites. $S_\text{kin, eff}$ is of the generic form \cite{Neuman2015,Glesaaen2016,Schoen2018}
\begin{align}
	S_\text{kin, eff} = \sum_x s_{\kappa^2}(U_x,U_{x+1}) + \sum_x s_{\kappa^4}(U_x,U_{x+1},U_{x+2})\;.
\end{align}
To perform the gauge integrals the Boltzmann weight is expanded to $\mathcal{O}(\kappa^4)$, 
\begin{align}
	Z =&\int[\dif{U}]\det Q_\text{stat}\prod_{\substack{x=1\\x\text{ mod } 2 = 0}}^{N_s}\sum_{r_{x-1}}\lambda_{r_{x-1}}\chi_{r_{x-1}}(U_{x-1})\chi_r(U_{x}^\dagger)\sum_{r_{x}}\lambda_{r_{x}}\chi_{r_{x}}(U_{x})\chi_r(U_{x+1}^\dagger)\notag\\
	&\times\left(1-s_{\kappa^2}(U_{x-1}, U_{x})-s_{\kappa^4}(U_{x-1}, U_{x},U_{x+1})+s_{\kappa^2}^2(U_{x-1}, U_{x})\right)\notag\\
	&\times\left(1-s_{\kappa^2}(U_{x}, U_{x+1})-s_{\kappa^4}(U_{x}, U_{x+1},U_{x+2})+s_{\kappa^2}^2(U_{x}, U_{x+1})\right)+\mathcal{O}(\kappa^6)\notag\\
	=&\int[\dif{U}]\det Q_\text{stat}\prod_{\substack{x=1\\x\text{ mod } 2 = 0}}^{N_x}\sum_{\substack{v_1^{},v_2^{},v_3^{},v_4^{}\\ r_1^{}, r_2^{}, r_3^{},r_4^{}}}h_{ v_1^{} v_2^{} v_3^{} v_4^{}}^{ r_1^{} r_2^{} r_3^{} r_4^{}}\chi_{r_1^{}}(U_{x-1})\chi_{r_2^{}}(U_{x}^\dagger)\chi_{r_3^{}}(U_{x})\chi_{r_4^{}}(U_{x+1}^\dagger)\notag\\
	&\times W_{v_1^{}}(U_{x-1})W_{v_2^{}}(U_{x})W_{v_3^{}}(U_{x+1})W_{v_4^{}}(U_{x+2})\;,
\end{align}
where we introduced the coupling tensor $\mathbf{h}$ and the indices $v_i$ carry the information of the powers and indices of the fractional Wilson loops \eqref{fractionalWilsonLoop}. After substituting $\mathbf{h}\to \mathbf{h}^{(n)}$ and integrating out every second pair $(U_{x-1}, U_{x})$ one finds the running couplings,
\begin{align}
	h_{ v_1^{} v_2^{} v_3^{} v_4^{}}^{  r_1^{}   r_2^{}  r_3^{}  r_4^{}(n+1)} =& \sum_{\substack{v_1^{\prime},v_2^{\prime},v_3^{\prime},v_4^{\prime}\\   r_1^{\prime},   r_2^{\prime},   r_3^{\prime},  r_4^{\prime}}}h_{ v_1^{} v_2^{} v_3^{\prime} v_4^{\prime}}^{  r_1^{}   r_2^{}  r_3^{}   r_4^{\prime}(n)}h_{ v_1^{\prime} v_2^{\prime} v_3^{} v_4^{}}^{  r_1^{\prime}   r_2^{\prime}   r_3^{\prime}  r_4^{}(n)}\left(\int\dif{U}\det Q_{\text{stat}}^{\text{loc}}(U)W_{v_1^{\prime}}(U)W_{v_3^{\prime}}(U)\chi_{  r_4^{\prime}}(U^\dagger)\chi_{ r_1^{\prime}}(U)\right)\notag\\
	&\times\left(\int\dif{U}\det Q_{\text{stat}}^{\text{loc}}(U)W_{v_2^{\prime}}(U)W_{v_4^{\prime}}(U)\chi_{  r_2^{\prime}}(U^\dagger)\chi_{  r_3^{\prime}}(U)\right)\label{kappa4run}\;.
\end{align}
\subsection{Evaluation of the renormalization scheme}
The expressions for the running couplings have a common structure. Any of them can be cast in a quadratic matrix recursion relation,
\begin{align}
	\mathbf{h}^{(n+1)} = \mathbf{h}^{(n)} \mathbf{g} \mathbf{h}^{(n)}\label{matRecRel}\;,
\end{align}
where $\mathbf{h}^{(n)}$ describes the running couplings, and the entries of $\mathbf{g}$ are given by the gauge integrals appearing in the recursion relations. It is quickly solved by
\begin{align}
	\boldsymbol{h}^{(n)} = \left(\boldsymbol{h}^{(0)}\boldsymbol{g}  \right)^{2^n -1}\boldsymbol{h}^{(0)}\label{recRelSol}\;.
\end{align}
If we assume that the spatial volume in lattice units is a power of two, $N_s = 2^{n_s}$, with periodic boundary conditions, the partition function can be calculated with the help of \eqref{recRelSol} and is given by
\begin{align}
	Z = \sum_{ij}h_{ij}^{(n_x)}g_{ji} = \tr \boldsymbol{h}^{(n_s)}\boldsymbol{g} =\tr \left[\left(\boldsymbol{h}^{(0)}\boldsymbol{g}\right)^{N_s}\right] = \tr \left[\left(\boldsymbol{g}\boldsymbol{h}^{(0)}\right)^{N_s}\right]\label{partgeneral}\;.
\end{align}
This solution allows us to take the thermodynamic limit. The remaining gauge integrals involve class functions only and can be performed analytically in the Polyakov gauge \cite{Nishida2004, Philipsen2019}.

In the following we are interested in the pressure $p$ and baryon density $n_B$ at non-zero chemical potential in the continuum for $N_f = 1$ and $N_c = 3$.  For static quarks representations with labels $0\le p,q\le 5$ are included in the gauge action. To set the scale we assume that the string tension $\sigma$ depends only weakly on $\kappa$, as in previous discussions on the 3d effective theory \cite{Glesaaen2016}. This allows us to use $\sigma$ in the pure gauge limit, $a^2\sigma = -\log u_f$ \cite{Huang1988}. Afterwards, the observables are fitted to a second-degree polynomial in the lattice spacing $a$. For the continuum extrapolation, $11$ equidistant points in the interval $\sqrt{\sigma}a\in\left[0.5,1\right]$ in the static quark limit and $\sqrt{\sigma}a\in\left[0.9,1.4\right]$ for non-static quarks are used \cite{ Neuman2015,Glesaaen2016}.
\begin{figure}[!t]
		\begin{subfigure}{.5\textwidth}
		\centering
		\includegraphics[width=\linewidth,page=1]{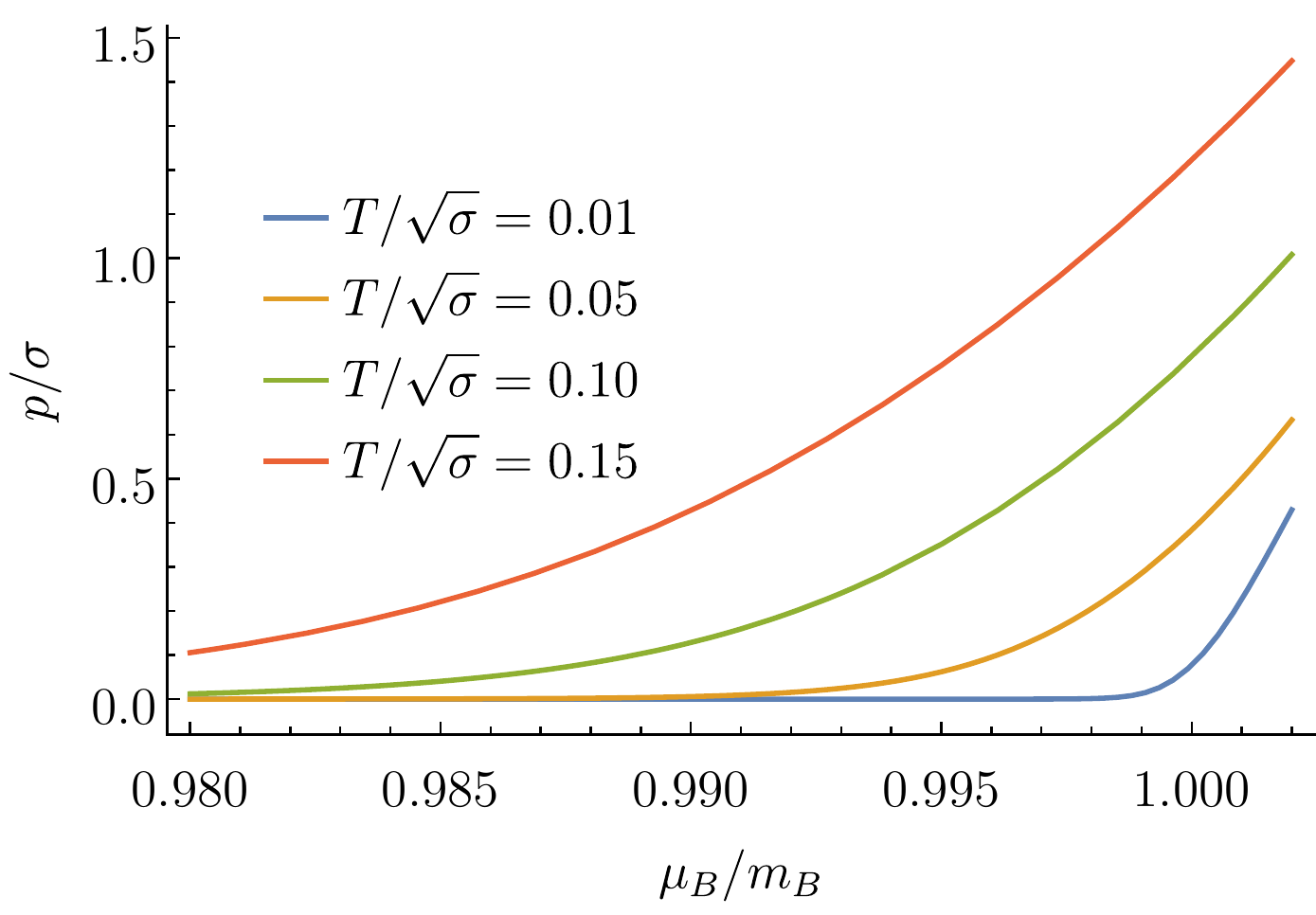}
		\caption{The continuum extrapolated pressure for varying \\ temperatures and static quarks.}
		\label{fig:PressureCont}
	\end{subfigure}%
	\begin{subfigure}{.5\textwidth}
		\centering
		\includegraphics[width=\linewidth,page=1]{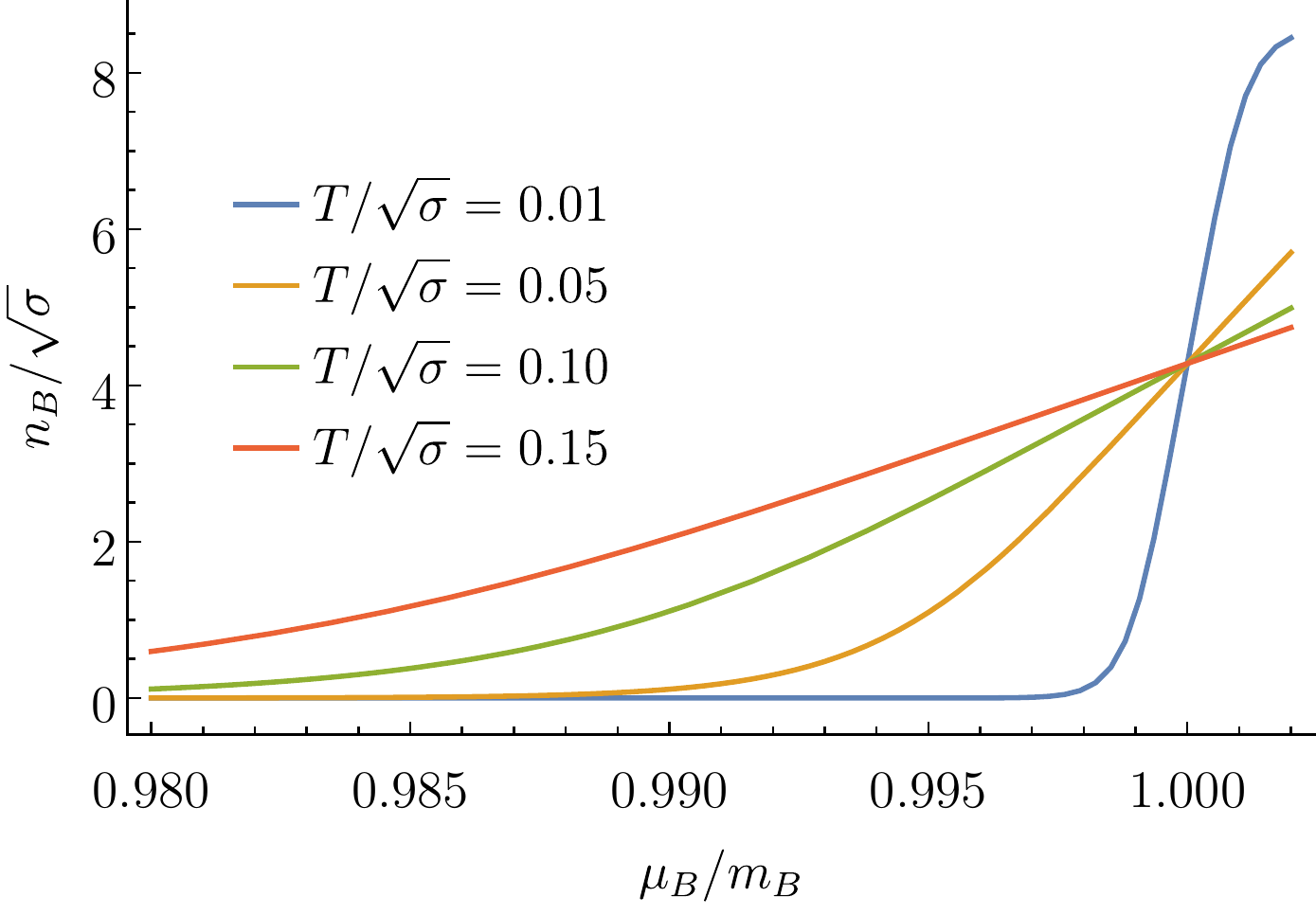}
		\caption{The continuum extrapolated baryon density for varying \\ temperatures and static quarks.}
		\label{fig:BaryonDensityCont}
	\end{subfigure}
\caption{Comparison of the pressure and baryon density vs. the baryon chemical potential in the static quark limit for $N_f = 1$, $m_B/\sqrt{\sigma} = 25$, $N_c = 3$ and varying temperatures.}
\label{fig:PressureStatic}
\end{figure}

In the static quark limit (figure \ref{fig:PressureStatic}) the hopping parameter $\kappa$ is chosen to satisfy $\kappa(\beta) < 10^{-2}$ for all values of beta. Thus, we set the baryon mass $\frac{m_B}{\sqrt{\sigma}} =  25$. Both plots show the same qualitative behavior as found in the effective theory in 3 spatial dimensions. The silver blaze property \cite{Cohen2003, Ipsen2012} can be observed, but no first-order transition at non-zero temperature is found \cite{Neuman2015,Glesaaen2016}. 

\begin{figure}[!t]
	\begin{subfigure}{.5\textwidth}
		\centering
		\includegraphics[width=\linewidth,page=1]{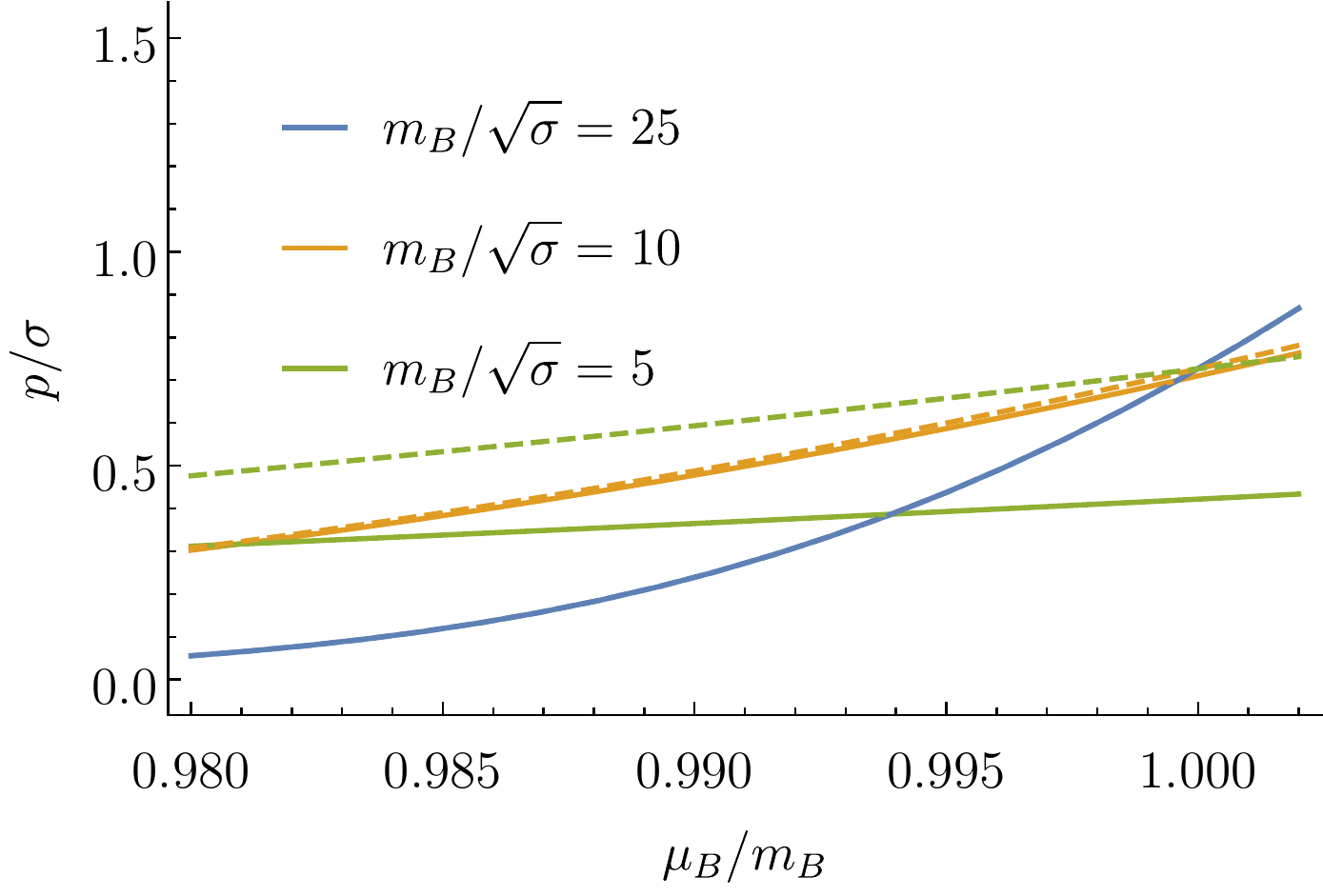}
		\caption{The pressure in the static quark limit (dashed lines) \\and to $\mathcal{O}(\kappa^2)$ (solid lines) for different baryon masses.}
		\label{fig:PressureHQKappa2}
	\end{subfigure}%
	\begin{subfigure}{.5\textwidth}
		\centering
		\includegraphics[width=\linewidth,page=1]{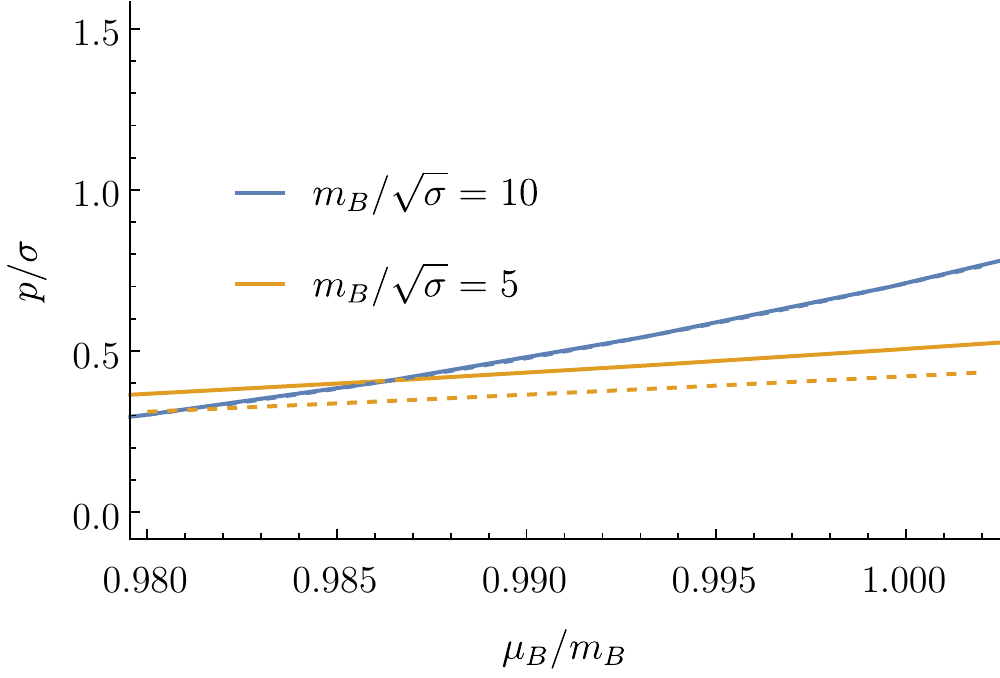}
		\caption{The pressure to $\mathcal{O}(\kappa^2)$ (dashes lines) and $\mathcal{O}(\kappa^4)$\\ (solid lines) for different baryon masses.}
		\label{fig:PressureKappa2Kappa4}
	\end{subfigure}
	\caption{The continuum extrapolated pressure vs. the baryon chemical potential for different orders in the heavy quark expansion for $N_f = 1$, $T/\sqrt{\sigma} = 0.15$, $N_c = 3$ and varying baryon masses.}
	\label{fig:PressureKappaComp}
\end{figure}

In figure \ref{fig:PressureKappaComp} the continuum extrapolated pressure is compared to different orders of the effective theories in $\kappa$ with  $T/\sqrt{\sigma} = 0.15$ and varying baryon masses. As shown in figure \ref{fig:PressureHQKappa2}, at high baryon masses ($m_B/\sqrt{\sigma} = 25$) leading order corrections are negligible. Approaching lower masses ($m_B/\sqrt{\sigma}  = 10$) first disagreements become apparent, which further increase when going to even smaller baryon masses ($m_B/\sqrt{\sigma}  = 5$). In contrast, differences between leading and next-to-leading order (figure \ref{fig:PressureKappa2Kappa4}) can only be observed for relatively light baryons ($m_B/\sqrt{\sigma}  = 5$).
\section{Coarse graining of the 2d effective theories}
In 2d renormalization group transformations are known to typically induce long range and multipoint interactions. To partially capture this behavior we allow next-to-nearest neighbor interactions, denoted by $[x,y]$. Analogously to the 2d Ising model \cite{Coldenfeld2018} the effective theories are renormalized by integrating out lattice sites in a checkerboard pattern. Note that next-to-nearest neighbor interactions start at $\mathcal{O}(u_f^{2N_\tau+2})$ \cite{ Langelage2011}.
\subsection{Pure gauge theory}
The partition function  with the recursion approach reads
\begin{align}
	Z = \int[\dif{U}]^{(n)}\prod_{\langle x,y \rangle^{(n)}}\Big(1+\lambda_f^{(n)} \left(L_x L_y^\dagger + L_y L_x^\dagger\right) \Big)\prod_{[ x,y ]^{(n)}}\Big(1+\lambda_{2,f}^{(n)} \left(L_x L_y^\dagger + L_y L_x^\dagger\right)\Big) + \mathcal{O}(u^{2N_\tau +4})\;.\label{2dpartstart}
\end{align}
To obtain the running couplings we consider a specific lattice site, say $x$, that we want to integrate out. Here we denote the product of the eight interaction terms between $x$ and its neighbors with $A(U_x)$. All terms of order $\mathcal{O}(u_f^{2N_\tau +4})$ in $A(U_x)$ are neglected and the remaining integrals can be solved with the orthogonality relation for irreducible characters. Afterwards, $A(U_x)$ can be rewritten into a product form \cite{Coldenfeld2018},
\begin{align}
	\int\dif{U_x}A(U_x) &= \left (1+\lambda _f^ {(n) 2}(L_ {x+{e_1} } L_ {x - {e_1}}^\dagger  +
	L_ {x - {e_1}} L_ {x+{e_1}}^\dagger)  \right) 
	\left (1+\lambda _f^ {(n) 2}(L_ {x-{e_1} } L_ {x - {e_2}}^\dagger  +
	L_ {x - {e_2}} L_ {x-{e_1}}^\dagger)  \right)  \notag\\
	&\quad\times \left (1+\lambda _f^ {(n) 2}(L_ {x+{e_2} } L_ {x - {e_1}}^\dagger  +
	L_ {x - {e_1}} L_ {x+{e_2}}^\dagger)  \right) 
	\left (1+\lambda _f^ {(n) 2}(L_ {x+{e_2} } L_ {x - {e_2}}^\dagger  +
	L_ {x - {e_2}} L_ {x+{e_2}}^\dagger)  \right) 
	\notag\\
	&\quad\times \left (1+\lambda _f^ {(n) 2}(L_ {x+{e_1} } L_ {x + {e_2}}^\dagger  +
	L_ {x + {e_2}} L_ {x+{e_1}}^\dagger)  \right) 
	\left (1+\lambda _f^ {(n) 2}(L_ {x+{e_1} } L_ {x - {e_2}}^\dagger  +
	L_ {x - {e_2}} L_ {x+{e_1}}^\dagger)  \right)  \notag\\
	&\quad+\mathcal{O}(u_f^{3N_\tau})\;.
\end{align}
After putting the products back into the partition function, the factors describe interactions between the remaining lattice sites at distances $|x-y| = 2a$ and $|x-y| = \sqrt{2}a$. The coupling strength of the former is given by $\lambda_f^{(n)2}$. For the latter the rotational symmetry of the system implies that the interaction appears twice. An additional contribution originates from the initial diagonal interactions between non-integrated lattice sites. After rotating the renormalized lattice by $\pi/4$ the running couplings read
\begin{align}
	\lambda_f^{(n+1)} &= 2\lambda _f^ {(n) 2}+\lambda _{2,f}^ {(n)}\quad\text{and}\quad	\lambda _{2,f}^ {(n+1)} = \lambda _f^ {(n) 2}\label{3Dpuregaugel1}\;.
\end{align}
\subsection{Static quark limit}
Now the static quark determinant is included into our description of the 2d effective theory. Like in the previous section we allow interactions over a  distance $\sqrt{2}a$. The renormalization scheme will be accurate to the order $\mathcal{O}(u_f^{vN_\tau} \kappa^{wN_\tau})$ with $v+w= 2$. For static quarks in 1d it was found that renormalization implies mixed terms between the different representations. Here this is applied to both the nearest neighbor and next-to-nearest neighbor interactions, respectively. This leads to
\begin{align}
	Z =& \int[\dif{U}]^{(n)}\det Q_\text{stat}^{(n)}\prod_{\langle x,y \rangle^{(n)}}\Big(1+\lambda_{1,1}^{(n)}(L_x^{}+L_y^{})+\lambda_{1,2}^{(n)}(L_x^\dagger+L_y^\dagger)+\lambda_{1,3}^{(n)}L_x^{} L_y^{}+\lambda_{1,4}^{(n)}L_x^\dagger L_y^\dagger\notag\\
	&+\lambda_{1,5}^{(n)}(L_x^{} L_y^\dagger + L_y^{} L_x^\dagger)\Big)
	\prod_{[ x,y ]^{(n)}}\Big(1+\lambda_{2,1}^{(n)}(L_x^{}+L_y^{})+\lambda_{2,2}^{(n)}(L_x^\dagger+L_y^\dagger)+\lambda_{2,3}^{(n)}L_x^{} L_y^{} +\lambda_{2,4}^{(n)}L_x^\dagger L_y^\dagger\notag\\
	&+\lambda_{2,5}^{(n)}(L_x^{} L_y^\dagger + L_y^{} L_x^\dagger)\Big) + \mathcal{O}(u^{2N_\tau +4})\label{eq:parthqapproach}\;.
\end{align}
 As in the previous section we integrate out the lattice sites in a checkerboard pattern. Before we do this, it is useful to define the gauge integral
\begin{align}
	 o(j,k):=\int\dif{U}\det Q_\text{stat}^\text{loc} L^jL^{\dagger k}  = \int\dif{U}L^jL^{\dagger k} + \mathcal{O}(\kappa^{N_\tau}) \label{hqint3D}\;.
\end{align}
The following steps are now analogous to the pure gauge limit. First, the product of the interaction terms of a single lattice site with all its neighbors is truncated at $\mathcal{O}(u_f^{vN_\tau} \kappa^{wN_\tau})$ with $v+w= 3$. Afterwards, it is brought back into product form and the renormalized lattice is rotated by $\pi/4$ \cite{Coldenfeld2018}. After pulling out a factor of $o(0,0)$ one finds the running couplings
\begin{align}
	\lambda_{1,5}^{(n+1)} = 2\frac{o(1,1)}{o(0,0)}\lambda_{1,5}^{(n)2}+\lambda_{2,5}^{(n)},\quad \lambda_{2,5}^{(n+1)} = \frac{o(1,1)}{o(0,0)}\lambda_{1,5}^{(n)2}\label{l15rec}\;.
\end{align}
Note, that the couplings $\lambda_{1,1}^{(n)}$, $\lambda_{1,2}^{(n)}$, $\lambda_{2,1}^{(n)}$ and $\lambda_{2,2}^{(n)}$ are always of order $\mathcal{O}(u_f^{vN_\tau} \kappa^{wN_\tau})$ with $v+w= 3$ and can therefore be disregarded. From the remaining running couplings one finds the non-trivial fixed point,
\begin{align}
	 \left(\lambda_{1,5}^{(n)},\lambda_{2,5}^{(n)}\right) = \left(\frac{o(0,0)}{3o(1,1)},\frac{o(0,0)}{9o(1,1)} \right) \label{critcouplingsq}\;.
\end{align}
To drive the analytical evaluation further we replace the initial next-to-nearest neighbor coupling $\lambda_{2,f}$ by $\lambda_f^2$. Then, the fixed point relation \eqref{critcouplingsq} is reduced to $\lambda_f(\beta_c) = \frac{o(0,0)}{3o(1,1)}\label{critl1rel}$,
which can be inverted numerically to find the critical inverse coupling $\beta_c$.
\subsection{The deconfinement transition}
 To test the effective theory in the pure gauge limit with $N_c = 3$ and $N_\tau = 2,\dots,5$ we compare $\beta_c$ with literature values from simulations of 2+1d Yang-Mills theory, taken from \cite{Liddle2008}, in table \ref{critbetasu3pg}. There, the solutions for $\beta_c$ are listed with their respective literature values $\beta_{c,\text{lit}}$ as well as their relative deviations $|\Delta\beta_c|/\beta_{c,\text{lit}} = |\beta_c-\beta_{c,\text{lit}}|/\beta_{c,\text{lit}}$. Our results in the pure gauge limit (table \ref{critbetasu3pg}) reproduce the literature values with an accuracy of about $12\%$.

\begin{table}[!h]
	\centering
	\begin{tabular}{|c|c|c|c|}
		\hline
		$N_\tau$& $\beta_c$ & $\beta_{c,\text{lit}}$ & $|\Delta\beta_c|/\beta_{c,\text{lit}}$ in $\%$ \\
		\hline
		2& 8.9267 &8.1489  & 9.5449\\
		\hline
		3& 12.6488 &11.3711  & 11.2365 \\
		\hline
		4&16.3067 &14.7170  & 10.8021 \\
		\hline
		5& 19.9532 &18.1310 & 10.0501 \\
		\hline
	\end{tabular}
	\captionof{table}{Critical inverse coupling $\beta_c$ for SU$(3)$ in the pure gauge limit. The literature values are taken from~  \cite{Liddle2008}.}
	\label{critbetasu3pg}
\end{table}

\begin{figure}[!h]
	
	\centering
	
	\includegraphics[width=0.75\linewidth]{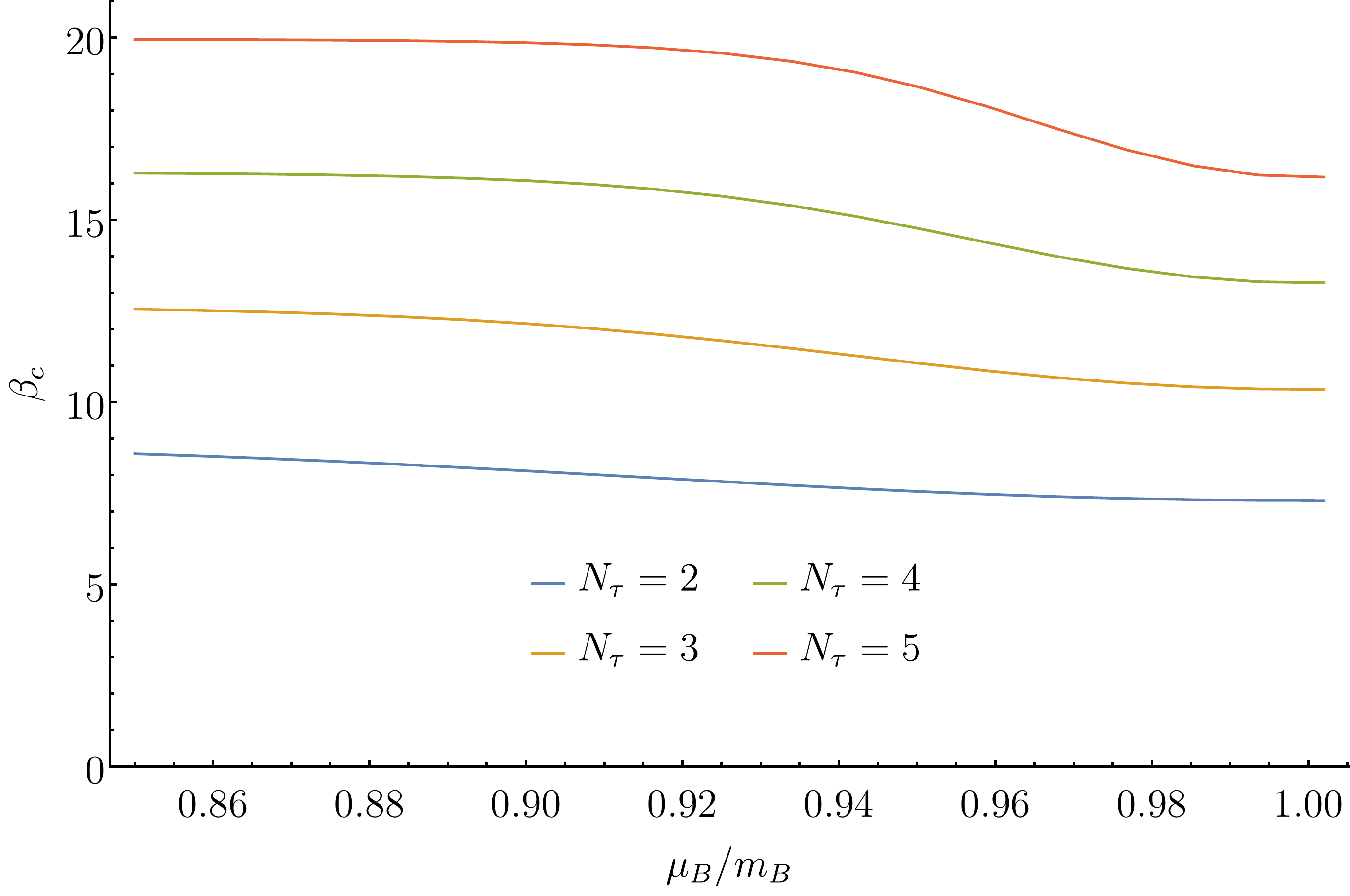}
	\captionof{figure}{The critical inverse coupling $\beta_c$ in the static quark limit for $N_c=3$ and $\kappa = 0.01$ vs. the baryon chemical potential $\mu_B/m_B$ for $N_\tau = 2,\dots, 5$.}
	\label{fig:Res2+1DStaticQuarks}
	
\end{figure}

In figure \ref{fig:Res2+1DStaticQuarks} it can be observed that an increasing chemical potential reduces the value of $\beta_c$. As this corresponds at fixed $N_\tau$ to a lower critical temperature, this is consistent with today's knowledge on the QCD phase diagram in 3+1d \cite{DEFORCRAND2002290}.

It is important to note that the renormalization scheme becomes less accurate at $\mu_B/m_B\approx 1 \approx h_1$ as the order of the transformation can be rewritten as $\mathcal{O}(u_f^{vN_\tau}\kappa^{wN_\tau}) = \mathcal{O}(u_f^{vN_\tau}h_1^{w_1}\bar h_1^{w_2})$ with $v+w = 2 = v + w_1 +w_2$.

\section{Conclusion}
Coarse graining techniques were applied to dimensionally reduced effective theories of 1+1d and 2+1d lattice QCD for heavy quarks. For the 1d effective theories the recursion relations of the running couplings were solved analytically using the transfer matrix. Afterwards, continuum extrapolation was performed numerically for some intensive observables. For the 2d effective theories coarse graining was applied in the pure gauge and static quark limit, respectively. The pure gauge critical couplings agree to about 12\% with simulation results of 2+1d Yang-Mills theory from the literature. 
\section*{Acknowledgments}
The authors acknowledge support by the Deutsche Forschungsgemeinschaft (DFG, German Research Foundation) through the CRC-TR 211 'Strong-interaction matter under extreme conditions'- project number 315477589 - TRR 211 and by the State of Hesse within the Research Cluster ELEMENTS (Project ID 500/10.006).

\bibliographystyle{JHEP}
\bibliography{library}
\end{document}